\RequirePackage{ifpdf}
\ifpdf 
\documentclass[pdftex]{sigma}
\else
\documentclass{sigma}
\fi

\DeclareMathOperator{\sech}{sech}
\DeclareMathOperator{\csch}{csch}
\DeclareMathOperator{\degen}{deg}

\begin{document}

\allowdisplaybreaks

\renewcommand{\PaperNumber}{067}

\FirstPageHeading

\numberwithin{equation}{section}

\newcommand{\field}[1]{\mathbb{#1}}
\newcommand{\Z}{\field{Z}}

\ShortArticleName{Position-Dependent Mass Schr\"odinger Equation}

\ArticleName{Quadratic Algebra Approach to an Exactly Solvable\\ Position-Dependent Mass
Schr\"odinger Equation\\ in Two Dimensions}

\Author{Christiane QUESNE}
\AuthorNameForHeading{C. Quesne}

\Address{Physique Nucl\'eaire Th\'eorique et Physique Math\'ematique,  Universit\'e Libre de
Bruxelles, \\ Campus de la Plaine CP229, Boulevard~du Triomphe, B-1050 Brussels,
Belgium}
\Email{\href{mailto:cquesne@ulb.ac.be}{cquesne@ulb.ac.be}}

\ArticleDates{Received March 30, 2007, in f\/inal form May
08, 2007; Published online May 17, 2007}

\Abstract{An exactly solvable position-dependent mass Schr\"odinger equation in two dimensions,
depicting a particle moving in a semi-inf\/inite layer, is re-examined in the light of recent theories
describing superintegrable two-dimensional systems with integrals of motion that are quadratic
functions of the momenta. To get the energy spectrum a quadratic algebra approach  is used together
with a realization in terms of deformed parafermionic oscillator operators. In this process, the
importance of supplementing algebraic considerations with a proper treatment of boundary conditions
for selecting physical wavefunctions is stressed. Some new results for matrix elements are derived.
This example emphasizes the interest of a quadratic algebra approach to position-dependent mass
Schr\"odinger equations.}


\Classification{81R12; 81R15}

\section{Introduction}

\looseness=1
Quantum mechanical systems with a position-dependent (ef\/fective) mass (PDM) have attracted a
lot of attention and inspired intense research activities during recent years. They are indeed very
useful in the study of many physical problems, such as electronic properties of
semiconductors~\cite{bastard} and quantum dots~\cite{serra}, nuclei~\cite{ring}, quantum
liquids~\cite{arias}, $^3$He clusters~\cite{barranco}, metal clusters~\cite{puente}, etc.

Looking for exact solutions of the Schr\"odinger equation with a PDM has become an interesting
research topic because such solutions may provide a conceptual understanding of some physical
phenomena, as well as a testing ground for some approximation schemes (for a list of references see,
e.g.,~\cite{cq06}). For such a purpose, use has been made of methods known in the constant-mass
case and extended to a PDM context, such as point canonical transformations~\cite{bhatta, natanzon,
levai89}, Lie algebraic methods~\cite{alhassid, wu, englefield, levai94}, as well as supersymmetric
quantum mechanical (SUSYQM) and shape-invariance techniques~\cite{cooper, bagchi}.

Although mostly one-dimensional equations have been considered up to now, several
works have recently paid attention to $d$-dimensional problems~\cite{cq06, chen, dong, mustafa06a,
mustafa06b, ju, gonul}. In \cite{cq06} (henceforth referred to as I and whose equations will be quoted
by their number preceded by I), we have analyzed $d$-dimensional PDM Schr\"odinger equations in the
framework of f\/irst-order intertwining operators and shown that with a pair $(H, H_1)$ of intertwined
Hamiltonians we can associate another pair $(R, R_1)$ of second-order partial dif\/ferential operators
related to the same intertwining operator and such that $H$ (resp.\ $H_1$) commutes with $R$ (resp.\
$R_1$). In the context of SUSYQM based on an sl(1/1) superalgebra, $R$ and $R_1$ can be interpreted
as SUSY partners, while $H$ and $H_1$ are related to the Casimir operator of a larger gl(1/1)
superalgebra.

In the same work, we have also applied our general theory to an explicit example, depicting a
particle moving in a two-dimensional semi-inf\/inite layer. This model may be of interest in the study
of quantum wires with an abrupt termination in an environment that can be modelled by a
dependence of the carrier ef\/fective mass on the position. It illustrates the inf\/luence of a~uniformity
breaking in a quantum channel on the production of bound states, as it was previously observed in
the case of a quantum dot or a bend~\cite{olendski, gudmunsson}.

From a theoretical viewpoint, our model has proved interesting too because it is solvable in two
dif\/ferent ways: by separation of variables in the corresponding Schr\"odinger equation or employing
SUSYQM and shape-invariance techniques. The former method relies upon the existence of an integral
of motion $L$, while, as above-mentioned, the latter is based on the use of~$R$. In other words, the
three second-order partial dif\/ferential operators $H$, $L$ and $R$ form a set of algebraically
independent integrals of motion, which means that the system is superintegrable.

Let us recall that in classical mechanics~\cite{goldstein}, an integrable system on a
$d$-dimensional mani\-fold is a system which has $d$ functionally independent (globally def\/ined)
integrals of motion in involution (including the Hamiltonian). Any system with more that $d$
functionally independent integrals of motion is called superintegrable. It is maximally superintegrable
if it admits the maximum number $2d-1$ of integrals of motion. The latter form a complete set so
that any other integral of motion can be expressed in terms of them. In particular, the Poisson
bracket of any two basic integrals, being again a constant of motion, can be written as a (in
general) nonlinear function of them. Such results can be extended to quantum
mechanics~\cite{dirac}, so that for quantum counterparts of maximally superintegrable systems we
get (in general) nonlinear associative algebras of algebraically independent observables, all of them
commuting with $H$.

The simplest case corresponds to the class of two-dimensional superintegrable systems with
integrals of motion that are linear and quadratic functions of the momenta. The study and
classif\/ication of such systems, dating back to the 19th century and revived in the
1960ties~\cite{fris, winternitz, makarov}, have recently been the subject of intense research
activities and substantial progress has been made in this area (see \cite{hietarinta, granovskii92a,
zhedanov, granovskii92b, granovskii92c, bonatsos, daska01, daska06a, daska06b, letourneau,
ranada97, ranada99, tempesta, kalnins97, kalnins99, kalnins05a, kalnins05b, kalnins06, kalnins07} and
references quoted therein). In particular, it has been shown that their integrals of motion generate a
quadratic Poisson algebra (in the classical case) or a quadratic associative algebra (in the quantum
one) with a Casimir of sixth degree in the momenta and the general form of these algebras has been
uncovered~\cite{daska01, kalnins05a, kalnins05b, kalnins06, kalnins07}. Algebras of this kind have many
similarities to the quadratic Racah algebra QR(3) (a special case of the quadratic Askey--Wilson
algebra QAW(3))~\cite{granovskii92a, zhedanov}. They actually coincide with QR(3) whenever one of
their parameters vanishes. The eigenvalues and eigenfunctions of the superintegrable system
Hamiltonian can be found from the f\/inite-dimensional irreducible representations of these algebras.
The latter can be determined by a ladder-operator method~\cite{granovskii92a, zhedanov,
granovskii92b, granovskii92c} or through a realization~\cite{bonatsos, daska01} in terms of
(generalized) deformed parafermionic operators~\cite{cq94}, which are a f\/inite-dimensional version
of deformed oscillator operators~\cite{daska91}.

Since our two-dimensional PDM model belongs to this class of superintegrable systems, it is
interesting to analyze it in the light of such topical and innovative theories. This is one of the
purposes of the present paper, which will therefore provide us with a third method for solving the
PDM Schr\"odinger equation. In such a process, we will insist on the necessity of supplementing
algebraic calculations with a proper treatment of the wavefunction boundary conditions imposed by
the physics of the problem -- a point that is not always highlighted enough.

Another purpose of this work is to stress the interest of a quadratic algebra approach to PDM
Schr\"odinger equations. If the presence of such an algebra was already noted before in a~one-dimensional example~\cite{roy}, this is indeed -- as far as the author knows -- the f\/irst
case where an algebra of this kind is used as a tool for solving a physical problem in a PDM
context.

This paper is organized as follows. In Section 2, the two-dimensional PDM model of I is brief\/ly
reviewed and some important comments on its mathematical structure are made in conjunction
with the physics of the problem. In Section 3, a quadratic algebra associated with such a~model is then
introduced and its classical limit is obtained. The f\/inite-dimensional irreducible representations of
the algebra are determined in Section 4. Finally, Section 5 contains the conclusion.

\section[Exactly solvable and superintegrable PDM model in a two-dimensional semi-infinite layer]{Exactly solvable and superintegrable PDM model\\ in a two-dimensional semi-inf\/inite layer}

In I we considered a particle moving in a two-dimensional semi-inf\/inite layer of width $\pi/q$,
parallel to the $x$-axis and with impenetrable barriers at the boundaries. The variables $x$, $y$ vary
in the domain
\begin{gather*}
  D: \qquad 0 < x < \infty, \qquad - \frac{\pi}{2q} < y < \frac{\pi}{2q},
\end{gather*}
and the wavefunctions have to satisfy the conditions
\begin{gather}
  \psi(0,y) = 0, \qquad \psi\left(x, \pm \frac{\pi}{2q}\right) = 0.  \label{eq:boundary2}
\end{gather}
The mass of the particle is $m(x) = m_0 M(x)$, where the dimensionless function $M(x)$ is given
by
\begin{gather}
  M(x) = \sech^2 qx.  \label{eq:mass}
\end{gather}
In units wherein $\hbar = 2 m_0 = 1$, the Hamiltonian of the model can be written as
\begin{gather}
  H^{(k)} = - \partial_x \frac{1}{M(x)} \partial_x - \partial_y \frac{1}{M(x)} \partial_y +
  V^{(k)}_{\rm eff}(x),  \label{eq:H}
\end{gather}
where we adopt the general form (I2.2) and
\begin{gather}
  V^{(k)}_{\rm eff}(x) = - q^2 \cosh^2 qx + q^2 k(k-1) \csch^2 qx  \label{eq:Veff}
\end{gather}
is an ef\/fective potential. This function includes some terms depending on the ambiguity
para\-me\-ters~\cite{vonroos}, which allow any ordering of the noncommutating momentum and PDM
operators (see equation~(I2.3)). In (\ref{eq:Veff}), the constant $k$ is assumed positive and we have set an
irrelevant additive constant $v_0$ to zero.

As shown in I, both the operators
\begin{gather*}
  L = - \partial_y^2
\end{gather*}
and
\begin{gather*}
  R^{(k)}  = \eta^{(k)\dagger} \eta^{(k)} \nonumber \\
  \phantom{R^{(k)}}{} = - \cosh^2 qx \sin^2 qy\, \partial^2_x + 2 \sinh qx \cosh qx \sin qy
       \cos qy\, \partial^2_{xy} - \sinh^2 qx \cos^2 qy\, \partial^2_y \nonumber \\
  \phantom{R^{(k)}=}{} + q \sinh qx \cosh qx (1 - 4 \sin^2 qy) \partial_x
       + q (1 + 4 \sinh^2 qx) \sin qy \cos qy \partial_y \nonumber \\
  \phantom{R^{(k)}=}{} + q^2 (\sinh^2 qx - \sin^2 qy - 3 \sinh^2 qx \sin^2 qy) - q^2 k (1 +
       \csch^2 qx \sin^2 qy) \nonumber \\
  \phantom{R^{(k)}=}{} + q^2 k^2 \csch^2 qx \sin^2 qy,
\end{gather*}
where{\samepage
\begin{gather*}
  \eta^{(k)\dagger}  = - \cosh qx \sin qy \,\partial_x + \sinh qx \cos qy\, \partial_y - q  \sinh qx
       \sin qy - qk \csch qx \sin qy, \\
  \eta^{(k)}  = \cosh qx \sin qy \,\partial_x - \sinh qx \cos qy\, \partial_y + q \sinh qx \sin qy - qk
       \csch qx \sin qy,
\end{gather*}
commute with} $H^{(k)}$, although not with one another. Hence one may diagonalize either
$H^{(k)}$ and $L$ or $H^{(k)}$ and $R^{(k)}$ simultaneously. This leads to two alternative bases
for the Hamiltonian eigenfunctions, corresponding to the eigenvalues
\begin{gather}
  E^{(k)}_N = q^2 (N+2) (N+2k+1), \qquad N=0, 1, 2, \ldots,  \label{eq:E}
\end{gather}
with degeneracies
\begin{gather}
  \degen(N) = \left[\frac{N}{2}\right] + 1,  \label{eq:deg}
\end{gather}
where $[N/2]$ stands for the integer part of $N/2$.

The f\/irst basis is obtained by separating the variables $x$, $y$ in the PDM Schr\"odinger equation
and its members, associated with the eigenvalues $(l+1)^2 q^2$ of $L$, read
\begin{gather}
  \psi^{(k)}_{n,l}(x,y) = \phi^{(k)}_{n,l}(x) \chi_l(y), \qquad  n, l = 0, 1, 2, \ldots, \label{eq:psi}
\end{gather}
with $N = 2n+l$,
\begin{gather}
  \phi^{(k)}_{n,l}(x) = {\cal N}^{(k)}_{n,l} (\tanh qx)^k (\sech qx)^{l+2}
       P^{\left(k-\frac{1}{2}, l+1\right)}_n(1 - 2 \tanh^2 qx),\nonumber\\
  \chi_l(y) = \begin{cases}
    \displaystyle  \sqrt{\frac{2q}{\pi}} \cos[(l+1)qy] & \text{for $l = 0, 2, 4, \ldots$}, \vspace{1mm}\\
   \displaystyle      \sqrt{\frac{2q}{\pi}} \sin[(l+1)qy] & \text{for $l = 1, 3, 5, \ldots$},
  \end{cases}   \label{eq:chi}
\end{gather}
and ${\cal N}^{(k)}_{n,l}$ a normalization constant given in equation~(I3.18).

The second basis, resulting from the intertwining relation
\begin{gather*}
  \eta^{(k)} H^{(k)} = H_1^{(k)} \eta^{(k)}, \qquad H_1^{(k)} = H^{(k+1)} + 2 q^2 k,
\end{gather*}
and its Hermitian conjugate, can be built by successive applications of operators of type $\eta^{(k)
\dagger}$,
\begin{gather}
  \Psi^{(k)}_{N,N_0}(x,y) = \bar{\cal N}^{(k)}_{N,N_0} \eta^{(k)\dagger}
  \eta^{(k+1)\dagger} \cdots \eta^{(k+\nu-1)\dagger} \Psi^{(k+\nu)}_{N_0,N_0}(x,y),
  \label{eq:Psi}
\end{gather}
on functions $\Psi^{(k+\nu)}_{N_0,N_0}(x,y)$, annihilated by $\eta^{(k+\nu)}$ and given in
Eqs.~(I3.28), (I3.32) and (I3.34). In (\ref{eq:Psi}), $N_0$ runs over 0, 2, 4,\ldots, $N$ or $N-1$,
according to whether $N$ is even or odd, while $\nu$, def\/ined by $\nu = N - N_0$, determines the
$R^{(k)}$ eigenvalue
\begin{gather}
  r^{(k)}_{\nu} = q^2 \nu (\nu + 2k), \qquad \nu=0, 1, 2, \ldots.  \label{eq:r}
\end{gather}
Although an explicit expression of the normalization coef\/f\/icient $\bar{\cal N}^{(k)}_{N,N_0}$ is
easily obtained (see equation~(I3.41)), this is not the case for $\Psi^{(k)}_{N,N_0}(x,y)$ (except for
some low values of $N$ and $N_0$), nor for the expansion of $\Psi^{(k)}_{N,N_0}(x,y)$ into the
f\/irst basis eigenfunctions $\psi^{(k)}_{n,l}(x,y)$, which is given by rather awkward formulas (see
equations~(I3.46), (I3.51), (I3.55) and (I3.56)).

Before proceeding to a quadratic algebra approach to the problem in Section 3, it is worth
making a few valuable observations, which were not included in I.

Mathematically speaking, the separable Schr\"odinger equation of our model admits four linearly
independent solutions obtained by combining the two independent solutions of the second-order
dif\/ferential equation in $x$ with those of the second-order dif\/ferential equation in $y$. Among
those four functions, only the combination $\psi^{(k)}_{n,l}(x,y)$, considered in (\ref{eq:psi}),
satisf\/ies all the boundary conditions and is normalizable on $D$. It is indeed clear that the alternative
solution to the dif\/ferential equation in $x$ is not normalizable, while that to the dif\/ferential
equation in $y$,
\begin{gather}
  \bar{\chi}_l(y) \propto \begin{cases}
      \sin[(l+1)qy] & \text{for $l = 0, 2, 4, \ldots$}, \\
      \cos[(l+1)qy] & \text{for $l = -1, 1, 3, 5, \ldots$},
  \end{cases}  \label{eq:chi-bar}
\end{gather}
violates the second condition in equation~(\ref{eq:boundary2}). Hence the three remaining combinations
provide unphysical functions.

Some mathematical considerations might also lead to another choice than $L$ and $R^{(k)}$ for
the basic integrals of motion complementing $H^{(k)}$. First of all, instead of $L$, one might
select the operator $p_y = - {\rm i} \partial_y$, which obviously satisf\/ies the condition
$[H^{(k)}, p_y] = 0$. This would result in a linear and a quadratic (in the momenta) integrals of
motion, generating a much simpler quadratic algebra than that to be considered in Section 3. It
should be realized, however, that the eigenfunctions $e^{{\rm i}my}$ ($m \in \mathbb{Z}$) of
$p_y$, being linear combinations of the physical and unphysical functions (\ref{eq:chi}) and
(\ref{eq:chi-bar}), are useless from a physical viewpoint. We are therefore forced to consider the
second-order operator $L$ instead of $p_y$.

Furthermore, it is straightforward to see that another pair of f\/irst-order dif\/ferential operators
\begin{gather}
  \bar{\eta}^{(k)\dagger}  = - \cosh qx \cos qy \,\partial_x - \sinh qx \sin qy\, \partial_y - q
       \sinh qx \cos qy - qk \csch qx \cos qy,  \label{eq:eta-bar-plus} \\
  \bar{\eta}^{(k)}  = \cosh qx \cos qy \,\partial_x + \sinh qx \sin qy\, \partial_y + q
       \sinh qx \cos qy - qk \csch qx \cos qy,  \label{eq:eta-bar}
\end{gather}
intertwines with $H^{(k)}$ and $H_1^{(k)}$, i.e., satisf\/ies the relation
\begin{gather}
  \bar{\eta}^{(k)} H^{(k)} = H_1^{(k)} \bar{\eta}^{(k)}, \qquad H_1^{(k)} = H^{(k+1)} + 2 q^2 k,
  \label{eq:inter-bar}
\end{gather}
and its Hermitian conjugate. Such operators correspond to the choice $a = c = g = 0$, $b = d = 1$
in equation~(I2.29).

As a consequence of (\ref{eq:inter-bar}), the operator
\begin{gather*}
  \bar{R}^{(k)}  = \bar{\eta}^{(k)\dagger} \bar{\eta}^{(k)} \\
\phantom{\bar{R}^{(k)}}{} = - \cosh^2 qx \cos^2 qy\, \partial^2_x - 2 \sinh qx \cosh qx \sin qy
       \cos qy\, \partial^2_{xy} - \sinh^2 qx \sin^2 qy\, \partial^2_y \\
\phantom{\bar{R}^{(k)}=}{} + q \sinh qx \cosh qx (1 - 4 \cos^2 qy) \partial_x
       - q (1 + 4 \sinh^2 qx) \sin qy \cos qy \partial_y \\
\phantom{\bar{R}^{(k)}=}{} + q^2 (\sinh^2 qx - \cos^2 qy - 3 \sinh^2 qx \cos^2 qy) - q^2 k (1 +
       \csch^2 qx \cos^2 qy) \\
\phantom{\bar{R}^{(k)}=}{}+ q^2 k^2 \csch^2 qx \cos^2 qy,
\end{gather*}
commutes with $H^{(k)}$ and is therefore another integral of motion. It can of course be
expressed in terms of $H^{(k)}$, $L$ and $R^{(k)}$, as it can be checked that
\begin{gather*}
  H^{(k)} = L + R^{(k)} + \bar{R}^{(k)} + 2 q^2 k.
\end{gather*}
However, we have now at our disposal three (dependent) integrals of motion $L$, $R^{(k)}$ and
$\bar{R}^{(k)}$ in addition to $H^{(k)}$, so that we may ask the following question: what is the
best choice for the basic integrals of motion from a physical viewpoint?

This problem is easily settled by noting that the zero modes of $\bar{\eta}^{(k)}$,
\begin{gather*}
  \bar{\omega}^{(k)}_s(x,y) = (\tanh qx)^k (\sech qx)^{s+1} (\sin qy)^s,
\end{gather*}
violate the second condition in equation~(\ref{eq:boundary2}) for any real value of $s$ and therefore
lead to unphysical functions. This contrasts with what happens for the zero modes
$\omega^{(k)}_s(x,y)$ of $\eta^{(k)}$, given in (I3.28), which are physical functions for $s > 0$
and can therefore be used to build the functions $\Psi^{(k)}_{N,N_0}(x,y)$ considered in
(\ref{eq:Psi}), as it was shown in (I3.32). We conclude that the physics of the model imposes
the choice of $L$ and $R^{(k)}$ as basic integrals of motion.

\section{Quadratic associative algebra and its classical limit}

It has been shown~\cite{daska01, kalnins06} that for any two-dimensional quantum superintegrable
system with integrals of motion $A$, $B$, which are second-order dif\/ferential operators, one can
construct a quadratic associative algebra generated by $A$, $B$, and their commutator $C$. This
operator is not independent of $A$, $B$, but since it is a third-order dif\/ferential operator, it cannot
be written as a~polynomial function of them. The general form of the quadratic algebra
commutation relations~is
\begin{gather}
  [A, B]  = C,  \label{eq:C1} \\
  [A, C]  = \alpha A^2 + \gamma \{A, B\} + \delta A + \epsilon B + \zeta,  \label{eq:C2} \\
  [B, C]  = a A^2 - \gamma B^2 - \alpha \{A, B\} + d A - \delta B + z.  \label{eq:C3}
\end{gather}
Here $\{A, B\} \equiv AB + BA$,
\begin{gather*}
  \delta  = \delta(H) = \delta_0 + \delta_1 H, \qquad \epsilon = \epsilon(H) = \epsilon_0 +
        \epsilon_1 H, \qquad \zeta = \zeta(H) = \zeta_0 + \zeta_1 H + \zeta_2 H^2, \\
  d  = d(H) = d_0 + d_1 H, \qquad z = z(H) = z_0 + z_1 H + z_2 H^2,
\end{gather*}
and $\alpha$, $\gamma$, $a$, $\delta_i$, $\epsilon_i$, $\zeta_i$, $d_i$, $z_i$ are some
constants. Note that it is the Jacobi identity $[A, [B, C]] = [B, [A, C]]$ that imposes some
relations between coef\/f\/icients in (\ref{eq:C2}) and (\ref{eq:C3}).

Such a quadratic algebra closes at level 6~\cite{kalnins06} or, in other words, it has a Casimir
operator which is a sixth-order dif\/ferential operator~\cite{daska01},
\begin{gather}
  K  = C^2 + \tfrac{2}{3} a A^3 - \tfrac{1}{3} \alpha \{A, A, B\} - \tfrac{1}{3} \gamma \{A, B,
        B\} + \left(\tfrac{2}{3} \alpha^2 + d + \tfrac{2}{3} a \gamma\right) A^2 \nonumber\\
\phantom{K=}{} + \left(\tfrac{1}{3} \alpha \gamma - \delta\right) \{A, B\}  + \left(\tfrac{2}{3} \gamma^2 -
        \epsilon\right) B^2 + \left(\tfrac{2}{3} \alpha \delta +
        \tfrac{1}{3} a \epsilon + \tfrac{1}{3} d \gamma + 2z\right) A \nonumber\\
\phantom{K=}{} + \left(- \tfrac{1}{3} \alpha \epsilon + \tfrac{2}{3} \gamma \delta - 2 \zeta\right)
        B + \tfrac{1}{3} \gamma z - \tfrac{1}{3} \alpha \zeta \nonumber\\
\phantom{K}{} = k_0 + k_1 H + k_2 H^2 + k_3 H^3,  \label{eq:K}
\end{gather}
where $k_i$ are some constants and $\{A, B, C\} \equiv ABC + ACB + BAC + BCA + CAB +
CBA$.

For our two-dimensional PDM model, described by the Hamiltonian def\/ined in equations (\ref{eq:mass})--(\ref{eq:Veff}), we shall take
\begin{gather}
  A = R, \qquad B = L,  \label{eq:A-B}
\end{gather}
where, for simplicity's sake, we dropped the superscript $(k)$ because no confusion can arise
outside the SUSYQM context.

To determine their commutation relations, it is worth noting f\/irst that their building blocks, the
f\/irst-order dif\/ferential operators $\partial_y$, $\eta^{\dagger}$ and $\eta$, generate another
quadratic algebra together with the other set of intertwining operators $\bar{\eta}^{\dagger}$,
$\bar{\eta}$, given in (\ref{eq:eta-bar-plus}) and (\ref{eq:eta-bar}). Their commutation relations
are indeed easily obtained as
\begin{alignat}{4}
 & [\partial_y, \eta]  = q \bar{\eta}, && [\partial_y, \bar{\eta}]  = - q \eta, &&
       [\eta, \bar{\eta}]  = q \partial_y, & \\  
 & [\eta, \eta^{\dagger}]  = 2 q^2 k (1 + \xi^2),\qquad & &  [\bar{\eta}, \bar{\eta}^{\dagger}]  =
       2 q^2 k (1 + \bar{\xi}^2),\qquad && [\eta, \bar{\eta}^{\dagger}]  = - q  \partial_y + 2 q^2 k  \xi
       \bar{\xi},  \label{eq:com-eta}&
\end{alignat}
and their Hermitian conjugates. In (\ref{eq:com-eta}), we have def\/ined
\begin{gather*}
  \xi = - (2qk)^{-1} (\eta + \eta^{\dagger}) = \csch qx \sin qy, \qquad \bar{\xi} = - (2qk)^{-1}
  (\bar{\eta} + \bar{\eta}^{\dagger}) = \csch qx \cos qy.
\end{gather*}
Interestingly, $\partial_y$, $\eta$ and $\bar{\eta}$ (as well as $\partial_y$, $\eta^{\dagger}$
and $\bar{\eta}^{\dagger}$) close an sl(2) subalgebra.

From these results, it is now straightforward to show that the operator $C$ in (\ref{eq:C1}) is
given by
\begin{gather*}
  C = q \{\partial_y, \eta^{\dagger} \bar{\eta} + \bar{\eta}^{\dagger} \eta\}
\end{gather*}
and that the coef\/f\/icients in (\ref{eq:C2}) and (\ref{eq:C3}) are
\begin{gather}
  \alpha  = \gamma = 8 q^2, \qquad \delta = 8 q^2 [q^2 (2k-1) - H], \qquad \epsilon = 16 q^4
       (k-1)(k+1), \nonumber\\
  \zeta  = 8 q^4 (k-1) (2 q^2 k - H), \qquad a = 0, \qquad d = 16 q^4, \qquad z = 8 q^4 (2
       q^2 k - H).
\label{eq:parameters}
\end{gather}
On inserting the latter in (\ref{eq:K}), we obtain for the value of the Casimir operator
\begin{gather*}
  K = - 4 q^4 [2q^2 (7k-6) - 3H] (2q^2 k - H).
\end{gather*}
It is worth noting that since $a=0$ in (\ref{eq:C3}), we actually have here an example of quadratic
Racah algebra QR(3)~\cite{granovskii92a}.

Before proceeding to a study of its f\/inite-dimensional irreducible representations in Section~4, it is
interesting to consider its classical limit. For such a purpose, since we have adopted units wherein
$\hbar = 2 m_0 = 1$, we have f\/irst to make a change of variables and of parameters restoring a
dependence on $\hbar$ (but keeping $2 m_0 = 1$ for simplicity's sake) before letting $\hbar$ go
to zero.

An appropriate transformation is
\begin{gather*}
  X = \hbar x, \qquad Y = \hbar y, \qquad P_X = - {\rm i} \hbar \partial_X, \qquad P_Y = - {\rm i}
  \hbar \partial_Y, \qquad Q = \frac{q}{\hbar}, \qquad K = \hbar k.
\end{gather*}
On performing it on the Hamiltonian given in equations~(\ref{eq:mass})--(\ref{eq:Veff}), we obtain
\begin{gather*}
  H = - \hbar^2 (\partial_X \cosh^2 QX \partial_X + \partial_Y \cosh^2 QX \partial_Y) - \hbar^2
  Q^2 \cosh^2 QX + Q^2 K(K - \hbar) \csch^2 QX,
\end{gather*}
yielding the classical Hamiltonian
\begin{gather*}
  H_{\rm c} = \lim_{\hbar \to 0} H = \cosh^2 QX (P_X^2 + P_Y^2) + Q^2 K^2 \csch^2 QX.
\end{gather*}
A similar procedure applied to the intertwining operators leads to
\begin{gather*}
  \eta_{\rm c}  = \lim_{\hbar \to 0} \eta = {\rm i} \cosh QX \sin QY P_X - {\rm i} \sinh
           QX \cos QY P_Y - QK \csch QX \sin QY,  \\
  \bar{\eta}_{\rm c}  = \lim_{\hbar \to 0} \bar{\eta} = {\rm i} \cosh QX \cos QY P_X +
           {\rm i} \sinh QX \sin QY P_Y - QK \csch QX \cos QY,
\end{gather*}
together with $\eta^*_{\rm c} = \lim\limits_{\hbar \to 0} \eta^{\dagger}$ and $\bar{\eta}^*_{\rm c} =
\lim\limits_{\hbar \to 0} \bar{\eta}^{\dagger}$, while the operators quadratic in the momenta give
rise to the functions
\begin{gather*}
  L_{\rm c}  = \lim_{\hbar \to 0} L = P_Y^2, \\
  R_{\rm c}  = \lim_{\hbar \to 0} R = \cosh^2 QX \sin^2 QY P_X^2 - 2 \sinh QX \cosh QX
       \sin QY \cos QY P_X P_Y \\
\phantom{R_{\rm c}  = \lim_{\hbar \to 0} R =}{} + \sinh^2 QX \cos^2 QY P_Y^2 + Q^2 K^2 \csch^2 QX \sin^2 QY, \\
  \bar{R}_{\rm c}  = \lim_{\hbar \to 0} \bar{R} = \cosh^2 QX \cos^2 QY P_X^2 + 2 \sinh QX
       \cosh QX \sin QY \cos QY P_X P_Y \\
\phantom{\bar{R}_{\rm c}  = \lim_{\hbar \to 0} \bar{R} =}{} + \sinh^2 QX \sin^2 QY P_Y^2 + Q^2 K^2 \csch^2 QX \cos^2 QY,
\end{gather*}
satisfying the relation
\begin{gather*}
  H_{\rm c} = L_{\rm c} + R_{\rm c} + \bar{R}_{\rm c}.
\end{gather*}

The quadratic associative algebra (\ref{eq:C1})--(\ref{eq:K}) is now changed into a quadratic
Poisson algebra, whose def\/ining relations can be determined either by taking the limit $
\lim\limits_{\hbar \to 0} ({\rm i} \hbar)^{-1} [O, O'] = \{O_{\rm c}, O'_{\rm c}\}_{\rm P}$ or by direct
calculation of the Poisson brackets $\{O_{\rm c}, O'_{\rm c}\}_{\rm P}$:
\begin{gather*}
  \{A_{\rm c}, B_{\rm c}\}_{\rm P}  = C_{\rm c}, \\
  \{A_{\rm c}, C_{\rm c}\}_{\rm P}  = \alpha_{\rm c} A_{\rm c}^2 + 2 \gamma_{\rm c}
        A_{\rm c} B_{\rm c} + \delta_{\rm c} A_{\rm c} + \epsilon_{\rm c} B_{\rm c} +
        \zeta_{\rm c}, \\
  \{B_{\rm c}, C_{\rm c}\}_{\rm P}  = a_{\rm c} A_{\rm c}^2 - \gamma_{\rm c}
        B_{\rm c}^2 - 2 \alpha_{\rm c} A_{\rm c} B_{\rm c} + d_{\rm c} A_{\rm c} - \delta_{\rm c}
        B_{\rm c} + z_{\rm c}.
\end{gather*}
Here
\begin{gather*}
  C_{\rm c} = \lim_{\hbar \to 0} \frac{C}{{\rm i} \hbar} = 2Q P_Y (\eta_{\rm c}^*
  \bar{\eta}_{\rm c} + \bar{\eta}_{\rm c}^* \eta_{\rm c})
\end{gather*}
and
\begin{gather*}
  \alpha_{\rm c} = \gamma_{\rm c} = - 8 Q^2, \qquad \delta_{\rm c} = 8 Q^2 H_{\rm c}, \qquad
  \epsilon_{\rm c} = - 16 Q^4 K^2, \qquad \zeta_{\rm c} = a_{\rm c} = d_{\rm c} = z_{\rm c} = 0.
\end{gather*}
Such a Poisson algebra has a vanishing Casimir:
\begin{gather*}
  K_{\rm c} = \lim_{\hbar \to 0} K = 0.
\end{gather*}

\section{Finite-dimensional irreducible representations\\ of the quadratic associative algebra}

The quadratic algebra (\ref{eq:C1})--(\ref{eq:K}) can be realized in terms of (generalized)
deformed oscillator operators $\cal N$, $b^{\dagger}$, $b$, satisfying the
relations~\cite{daska91}
\begin{gather*}
  [{\cal N}, b^{\dagger}] = b^{\dagger}, \qquad [{\cal N}, b] = - b, \qquad b^{\dagger} b =
  \Phi({\cal N}), \qquad b b^{\dagger} = \Phi({\cal N}+1),
\end{gather*}
where the structure function $\Phi(x)$ is a `well-behaved' real function such that
\begin{gather}
  \Phi(0) = 0, \qquad \Phi(x) > 0 \quad {\rm for} \quad x > 0.  \label{eq:Phi-C1}
\end{gather}
This deformed oscillator algebra has a Fock-type representation, whose basis states $|m\rangle$,
$m=0$, $1,2,\ldots$,\footnote{We adopt here the unusual notation $|m\rangle$ in order to avoid
confusion between the number of deformed bosons and the quantum number $n$ introduced in
(\ref{eq:psi}).} fulf\/il the relations
\begin{gather}
\begin{split}
  & {\cal N} |m\rangle = m |m\rangle, \\
  & b^{\dagger} |m\rangle = \sqrt{\Phi(m+1)}\, |m+1\rangle, \qquad m = 0, 1, 2, \ldots, \\
  & b |0\rangle = 0, \\
  & b |m\rangle = \sqrt{\Phi(m)}\, |m-1\rangle, \qquad m = 1, 2, \ldots.
\end{split}  \label{eq:Fock}
\end{gather}

We shall be more specif\/ically interested here in a subclass of deformed oscillator operators, which
have a ($p+1$)-dimensional Fock space, spanned by $|p, m\rangle \equiv |m\rangle$, $m=0,
1,\ldots, p$, due to the following property
\begin{gather}
  \Phi(p+1) = 0  \label{eq:Phi-C2}
\end{gather}
of the structure function, implying that
\begin{gather*}
  (b^{\dagger})^{p+1} = b^{p+1} = 0.
\end{gather*}
These are so-called (generalized) deformed parafermionic oscillator operators of order
$p$~\cite{cq94}. The general form of their structure function is given by
\begin{gather*}
  \Phi(x) = x (p+1-x) (a_0 + a_1 x + a_2 x^2 + \cdots + a_{p-1} x^{p-1}),
\end{gather*}
where $a_0, a_1,\ldots, a_{p-1}$ may be any real constants such that the second condition
in (\ref{eq:Phi-C1}) is satisf\/ied for $x=1, 2,\ldots, p$.

A realization of the quadratic algebra (\ref{eq:C1})--(\ref{eq:K}) in terms of deformed oscillator
operators $\cal N$, $b^{\dagger}$, $b$ reads~\cite{daska01}
\begin{gather}
  A  = A({\cal N}), \label{eq:A-para} \\
  B  = \sigma({\cal N}) + b^{\dagger} \rho({\cal N}) + \rho({\cal N}) b,  \label{eq:B-para}
\end{gather}
where $A({\cal N})$, $\sigma({\cal N})$ and $\rho({\cal N})$ are some functions of $\cal N$,
which, in the $\gamma \ne 0$ case, are given by
\begin{gather}
  A({\cal N})  = \frac{\gamma}{2} \left[({\cal N}+u)^2 - \frac{1}{4} -
         \frac{\epsilon}{\gamma^2}\right], \\
  \sigma({\cal N})  = - \frac{\alpha}{4} \left[({\cal N}+u)^2 - \frac{1}{4}\right] +
         \frac{\alpha\epsilon - \gamma\delta}{2 \gamma^2} - \frac{\alpha\epsilon^2 - 2 \gamma
         \delta\epsilon + 4 \gamma^2 \zeta}{4 \gamma^4} \frac{1}{({\cal N}+u)^2 - \frac{1}{4}},
         \label{eq:sigma} \\
  \rho^2({\cal N})  = \frac{1}{3 \cdot 2^{12} \gamma^8 ({\cal N}+u) ({\cal N}+u+1) [2 ({\cal
         N}+u)+1]^2},  \label{eq:rho}
\end{gather}
with the structure function
\begin{gather}
  \Phi(x)  = - 3072 \gamma^6 K [2 ({\cal N}+u)-1]^2 \nonumber\\
\phantom{\Phi(x)  =}{} - 48 \gamma^6 (\alpha^2 \epsilon - \alpha\gamma\delta + a\gamma\epsilon -
       d\gamma^2) [2 ({\cal N}+u)-3] [2 ({\cal N}+u)-1]^4 [2 ({\cal N}+u)+1] \nonumber\\
\phantom{\Phi(x)  =}{} + \gamma^8 (3\alpha^2 + 4a\gamma) [2 ({\cal N}+u)-3]^2 [2 ({\cal N}+u)-1]^4
       [2 ({\cal N}+u)+1]^2 \nonumber\\
\phantom{\Phi(x)  =}{} + 768 (\alpha\epsilon^2 - 2\gamma\delta\epsilon + 4\gamma^2 \zeta)^2 \nonumber\\
\phantom{\Phi(x)  =}{} + 32 \gamma^4 (3\alpha^2 \epsilon^2 - 6\alpha\gamma\delta\epsilon +
       2a\gamma\epsilon^2 + 2\gamma^2 \delta^2 - 4d\gamma^2 \epsilon + 8\gamma^3 z +
       4\alpha\gamma^2 \zeta) \nonumber\\
\phantom{\Phi(x)  =}{} \times [2 ({\cal N}+u)-1]^2 [12({\cal N}+u)^2 - 12({\cal N}+u) - 1] \nonumber\\
\phantom{\Phi(x)  =}{} - 256 \gamma^2 (3\alpha^2 \epsilon^3 - 9\alpha\gamma\delta\epsilon^2 +
       a\gamma\epsilon^3 + 6\gamma^2 \delta^2 \epsilon - 3d\gamma^2 \epsilon^2 +
       2\gamma^4 \delta^2 + 2d\gamma^4 \epsilon + 12\gamma^3 \epsilon z \nonumber\\
\phantom{\Phi(x)  =}{} - 4\gamma^5 z + 12\alpha\gamma^2 \epsilon\zeta - 12\gamma^3 \delta
       \zeta + 4\alpha\gamma^4 \zeta) [2 ({\cal N}+u)-1]^2.  \label{eq:Phi-gen}
\end{gather}
These functions depend upon two (so far undetermined) constants, $u$ and the eigenvalue of the
Casimir operator $K$ (which we denote by the same symbol).

Such a realization is convenient to determine the representations of the quadratic algebra in a basis
wherein the generator $A$ is diagonal together with $K$ (or, equivalently, $H$) because the
former is already diagonal with eigenvalues $A(m)$. The ($p+1$)-dimensional representations,
associated with ($p+1$)-fold degenerate energy levels, correspond to the restriction to deformed
parafermionic operators of order $p$~\cite{daska01}. The f\/irst condition in (\ref{eq:Phi-C1}) can
then be used with equation~(\ref{eq:Phi-C2}) to compute $u$ and $K$ (or $E$) in terms of $p$ and of the
Hamiltonian parameters. A choice is then made between the various solutions that emerge from the
calculations by imposing the second restriction in (\ref{eq:Phi-C1}) for $x=1, 2,\ldots, p$.

In the present case, for the set of parameters (\ref{eq:parameters}), the complicated structure
function (\ref{eq:Phi-gen}) drastically simplif\/ies to yield the factorized expression
\begin{gather*}
  \Phi(x)  = 3 \cdot 2^{30} q^{20} (2x+2u+k-1) (2x+2u+k-2) (2x+2u-k) (2x+2u-k-1) \\
\phantom{\Phi(x)  =}{} \times \left(2x+2u- \tfrac{1}{2} + \Delta\right) \left(2x+2u - \tfrac{3}{2} +
       \Delta\right) \left(2x+2u- \tfrac{1}{2} - \Delta\right) \left(2x+2u- \tfrac{3}{2} - \Delta\right),
\end{gather*}
where
\begin{gather*}
  \Delta = \sqrt{\left(k - \frac{1}{2}\right)^2 + \frac{E}{q^2}}.
\end{gather*}
Furthermore, the eigenvalues of the operator $A$ become
\begin{gather*}
  A(m) = q^2 (2m+2u-k) (2m+2u+k).
\end{gather*}
Since $A=R$ is a positive-def\/inite operator, only values of $u$ such that $A(m) \ge 0$ for $m=0$,
$1,\ldots, p$ should be retained.

On taking this into account, the f\/irst condition in (\ref{eq:Phi-C1}) can be satisf\/ied by choosing
either $u = k/2$ or $u = (k+1)/2$, yielding
\begin{gather}
  A(m) = 4 q^2 m(m+k)  \label{eq:A-1}
\end{gather}
or
\begin{gather}
  A(m) = 4 q^2 \left(m + \tfrac{1}{2}\right) \left(m + k + \tfrac{1}{2}\right),  \label{eq:A-2}
\end{gather}
respectively. For $u = k/2$, equation~(\ref{eq:Phi-C2}), together with the second condition in
(\ref{eq:Phi-C1}), can be fulf\/illed in two dif\/ferent ways corresponding to $\Delta = 2p + k + 1
\pm \frac{1}{2}$ or
\begin{gather}
  E = q^2 \left(2p + \tfrac{3}{2} \pm \tfrac{1}{2}\right) \left(2p + 2k + \tfrac{1}{2} \pm
  \tfrac{1}{2}\right).  \label{eq:E-1}
 \end{gather}
The resulting structure function reads
\begin{gather}
  \Phi(x)  = 3 \cdot 2^{38} q^{20} x (p+1-x) \left(x - \tfrac{1}{2}\right) \left(p+1 \pm
       \tfrac{1}{2} - x\right) \left(x+k - \tfrac{1}{2}\right) (x+k-1) \nonumber\\
\phantom{\Phi(x)  =}{} \times \left(x+p+k + \tfrac{1}{4} \pm \tfrac{1}{4}\right) \left(x+p+k -
       \tfrac{1}{4} \pm \tfrac{1}{4}\right).  \label{eq:Phi-1}
\end{gather}
Similarly, for $u = (k+1)/2$, we obtain
\begin{gather}
  E = q^2 \left(2p + \tfrac{5}{2} \pm \tfrac{1}{2}\right) \left(2p + 2k + \tfrac{3}{2} \pm
  \tfrac{1}{2}\right)  \label{eq:E-2}
 \end{gather}
and
\begin{gather}
  \Phi(x)  = 3 \cdot 2^{38} q^{20} x (p+1-x) \left(x + \tfrac{1}{2}\right) \left(p+1 \pm
       \tfrac{1}{2} - x\right) (x+k) \left(x+k - \tfrac{1}{2}\right)  \nonumber\\
\phantom{\Phi(x)  =}{} \times \left(x+p+k + \tfrac{5}{4} \pm \tfrac{1}{4}\right) \left(x+p+k +
       \tfrac{3}{4} \pm \tfrac{1}{4}\right).  \label{eq:Phi-2}
\end{gather}

Our quadratic algebra approach has therefore provided us with a purely algebraic derivation of the
eigenvalues of $H$ and $R$ in a basis wherein they are simultaneously diagonal. It now remains to
see to which eigenvalues we can associate physical wavefunctions, i.e., normalizable functions
satisfying equation~(\ref{eq:boundary2}). This will imply a correspondence between $|p, m\rangle$ and
the functions $\Psi_{N, N-\nu}(x,y)$, def\/ined in (\ref{eq:Psi}).

On comparing $A(m)$ to the known (physical) eigenvalues $r_{\nu}$ of $R$, given in (\ref{eq:r}),
we note that the f\/irst choice (\ref{eq:A-1}) for $A(m)$ corresponds to even $\nu = 2m$ (hence
to even $N$), while the second choice (\ref{eq:A-2}) is associated with odd $\nu = 2m+1$ (hence
with odd $N$). Appropriate values of $p$ leading to the level degeneracies (\ref{eq:deg}) are
therefore $p = N/2$ and $p = (N-1)/2$, respectively. With this identif\/ication, both
equations~(\ref{eq:E-1}) and (\ref{eq:E-2}) yield the same result
\begin{gather}
  E = q^2 \left(N + \tfrac{3}{2} \pm \tfrac{1}{2}\right) \left(N + 2k + \tfrac{1}{2} \pm
  \tfrac{1}{2}\right).  \label{eq:E-12}
\end{gather}
Comparison with (\ref{eq:E}) shows that only the upper sign choice in (\ref{eq:E-12}) leads to
physical wavefunctions $\Psi_{N, N-\nu}(x,y)$.

Restricting ourselves to such a choice, we can now rewrite all the results obtained in this section
in terms of $N$ and $\nu$ instead of $p$ and $m$. In particular, the two expressions
(\ref{eq:Phi-1}) and~(\ref{eq:Phi-2}) for the structure function can be recast in a single form
$\Phi(m) \to \Phi_{\nu}$, where{\samepage
\begin{gather}
  \Phi_{\nu}  = 3 \cdot 2^{30} q^{20} \nu (\nu-1) (\nu+2k-1) (\nu+2k-2) (N+\nu+2k) (N+\nu+2k
  +1) \nonumber\\
\phantom{\Phi_{\nu}  =}{} \times (N-\nu+2) (N-\nu+3).  \label{eq:Phi-nu}
\end{gather}}

More importantly, our quadratic algebra analysis provides us with an entirely new result, namely
the matrix elements of the integral of motion $L$ in the basis wherein $H$ and $R$ are
simultaneously diagonal. On using indeed the correspondence $|p, m\rangle \to \Psi_{N, N-\nu}$,
as well as the results in equations~(\ref{eq:Fock}), (\ref{eq:B-para}), (\ref{eq:sigma}), (\ref{eq:rho})
and (\ref{eq:Phi-nu}), we obtain
\begin{gather}
  L \Psi_{N, N-\nu} = \sigma_{\nu} \Psi_{N, N-\nu} + \tau_{\nu} \Psi_{N, N-\nu+2} +
  \tau_{\nu+2} \Psi_{N, N-\nu-2},  \label{eq:L-me}
\end{gather}
where we have reset $\sigma(m) \to \sigma_{\nu}$, $\rho(m) \to \rho_{\nu}$ and def\/ined
$\tau_{\nu} = s_{\nu} \rho_{\nu-2} \sqrt{\Phi_{\nu}}$. The explicit form of the coef\/f\/icients
 on the right-hand side of (\ref{eq:L-me}) is given by
\begin{gather}
  \sigma_{\nu}  = \frac{q^2}{2(\nu+k-1)(\nu+k+1)} \{- (\nu+k-1)^2 (\nu+k+1)^2\nonumber \\
\phantom{\sigma_{\nu}  =}{} + [N^2 + (2k+3)N + 2k^2 + 2k +1] (\nu+k-1) (\nu+k+1) \nonumber\\
\phantom{\sigma_{\nu}  =}{} - k(k-1)(N+k+1)(N+k+2)\},
\\
  \tau_{\nu}^2  = \frac{q^4}{16(\nu+k-2) (\nu+k-1)^2 (\nu+k)} \nu (\nu-1) (\nu+2k-1) (\nu+2k-2) \nonumber\\
\phantom{\tau_{\nu}^2  =}{} \times (N-\nu+2) (N-\nu+3) (N+\nu+2k) (N+\nu+2k+1).  \label{eq:tau}
\end{gather}
Note that $\tau_{\nu}$ is determined up to some phase factor $s_{\nu}$ depending on the
convention adopted for the relative phases of $\Psi_{N, N-\nu}$ and $\Psi_{N, N-\nu + 2}$.

For $N=4$, for instance, $\nu$ runs over 0, 2, 4, so that equations~(\ref{eq:L-me})--(\ref{eq:tau})
become
\begin{gather*}
  L \Psi_{4,0} = \frac{q^2}{k+3} \Biggl[(13k+21) \Psi_{4,0} + 3s_4 \sqrt{\frac{2(k+1)(2k+3)
       (2k+9)}{k+2}} \Psi_{4,2}\Biggr], \\
       L \Psi_{4,2}  = q^2 \Biggl[\frac{3s_4}{k+3} \sqrt{\frac{2(k+1)(2k+3)(2k+9)}{k+2}}
           \Psi_{4,0} + \frac{17k^2+76k+39}{(k+1)(k+3)} \Psi_{4,2} \\
\phantom{L \Psi_{4,2}  =}{}+ \frac{s_2}{k+1} \sqrt{\frac{10(k+3)(2k+1)(2k+7)}{k+2}} \Psi_{4,4}\Biggr],
\\
  L \Psi_{4,4} = \frac{q^2}{k+1} \Biggl[s_2 \sqrt{\frac{10(k+3)(2k+1)(2k+7)}{k+2}}
       \Psi_{4,2} + 5(k+3) \Psi_{4,4}\Biggr].
\end{gather*}
As a check, these results can be compared with those derived from the action of $L$ on the
expansions of $\Psi_{4,0}$, $\Psi_{4,2}$ and $\Psi_{4,4}$ in terms of the f\/irst basis
eigenfunctions $\psi_{0,4}$, $\psi_{1,2}$ and $\psi_{2,0}$ (see, e.g., equations~(I3.61) and (I3.49) for
$\Psi_{4,0}$ and $\Psi_{4,4}$, respectively). This leads to the phase factors $s_2 = s_4 = -1$.

To conclude, it is worth mentioning that had we made the opposite choice in equation~(\ref{eq:A-B}),
i.e., $A=L$ and $B=R$, we would not have been able to use the deformed parafermionic realization
(\ref{eq:A-para}), (\ref{eq:B-para}) to determine the energy spectrum. The counterpart of the
parafermionic vacuum state would indeed have been a function annihilated by $L$ and therefore
constructed from the unphysical function $\bar{\chi}_{-1}(y)$ of equation~(\ref{eq:chi-bar}).

\section{Conclusion}

In this paper, we have revisited the exactly solvable PDM model in a two-dimensional semi-inf\/inite
layer introduced in I. Here we have taken advantage of its superintegrability with two integrals of
motion $L$ and $R$ that are quadratic in the momenta to propose a third method of solution in the
line of some recent analyses of such problems.

We have f\/irst determined the explicit form of the quadratic associative algebra generated by~$L$,~$R$ and their commutator. We have shown that it is a quadratic Racah algebra QR(3) and that its
Casimir operator $K$ is a second-degree polynomial in $H$. We have also obtained the quadratic
Poisson algebra arising in the classical limit.

We have then studied the f\/inite-dimensional irreducible representations of our algebra in a~basis
wherein $K$ (or $H$) and $R$ are diagonal. For such a purpose, we have used a simple procedure,
proposed in \cite{daska01}, consisting in mapping this basis onto deformed parafermionic oscillator
states of order $p$. Among the results so obtained for the energy spectrum, we have selected
those with which physical wavefunctions can be associated. This has illustrated once again the
well-known fact that in quantum mechanics the physics is determined not only by algebraic
properties of operators, but also by the boundary conditions imposed on wavefunctions. Our
analysis has provided us with an interesting new result, not obtainable in general form in the
SUSYQM approach of I, namely the matrix elements of $L$ in the basis wherein $H$ and $R$ are
simultaneously diagonal.

As f\/inal points, it is worth observing that the approaches followed here are not the only ones available.
First, one could have used a gauge transformation to relate equation (\ref{eq:H}) to a well-known superintegrable system in a Darboux space (\cite{daska06b, kalnins07} and references quoted therein). Second, the irreducible representations of QR(3) could have been constructed by the ladder-operator method
employed in~\cite{granovskii92a, zhedanov, granovskii92b, granovskii92c}. This would have allowed us to express the transformation matrix elements between the bases $\psi^{(k)}_{n,l}$ and $\Psi^{(k)}_{N,N_0}$
(denoted by $Z^{(k)}_{N_0;n,l}$ in I) in terms of Racah--Wilson polynomials.

\pdfbookmark[1]{References}{ref}
\LastPageEnding


\begin{thebibliography}{99}

\footnotesize\itemsep=0pt

\bibitem{bastard} Bastard G., Wave mechanics applied to semiconductor heterostructures, Editions
de Physique, Les Ulis, 1988.

\bibitem{serra} Serra L., Lipparini E., Spin response of unpolarized quantum dots, {\it Europhys.\ Lett.}
{\bf 40} (1997), 667--672.

\bibitem{ring} Ring P., Schuck P., The nuclear many body problem, Springer, New York, 1980.

\bibitem{arias} Arias de Saavedra F., Boronat J., Polls A., Fabrocini A., Ef\/fective mass of one ${}^4$He
atom in liquid ${}^3$He, {\it Phys.\ Rev.\ B} {\bf 50} (1994), 4248--4251, \href{http://arxiv.org/abs/cond-mat/9403075}{cond-mat/9403075}.

\bibitem{barranco} Barranco M., Pi M., Gatica S.M., Hern\'andez E.S., Navarro J., Structure and
energetics of mixed ${}^4$He-${}^3$He drops, {\it Phys.\ Rev.\ B} {\bf 56} (1997), 8997--9003.

\bibitem{puente} Puente A., Serra Ll., Casas M., Dipole excitation of Na clusters with a non-local
energy density functional, {\it Z.\ Phys.\ D} {\bf 31} (1994), 283--286.

\bibitem{cq06} Quesne C., First-order intertwining operators and position-dependent mass
Schr\"odinger equations in $d$ dimensions, {\it Ann. Physics} {\bf 321} (2006), 1221--1239,
\href{http://arxiv.org/abs/quant-ph/0508216}{quant-ph/0508216}.

\bibitem{bhatta} Bhattacharjie A., Sudarshan E.C.G., A class of solvable potentials, {\it Nuovo
Cimento} {\bf 25} (1962), 864--879.

\bibitem{natanzon} Natanzon G.A., General properties of potentials for which the Schr\"odinger
equation can be solved by means of hypergeometric functions, {\it Theoret. and Math.\ Phys.} {\bf 38}
(1979), 146--153.

\bibitem{levai89} L\'evai G., A search for shape-invariant solvable potentials, {\it J.\ Phys.\ A: Math.\
Gen.} {\bf 22} (1989), 689--702.

\bibitem{alhassid} Alhassid Y., G\"ursey F., Iachello F., Group theory approach to scattering. II. The
Euclidean connection, {\it Ann. Physics} {\bf 167} (1986), 181--200.

\bibitem{wu} Wu J., Alhassid Y., The potential group approach and hypergeometric dif\/ferential
equations, {\it J.\ Math.\ Phys.} {\bf 31} (1990), 557--562.

\bibitem{englefield} Englef\/ield M.J., Quesne C., Dynamical potential algebras for Gendenshtein and
Morse potentials, {\it J.~Phys.~A: Math.\ Gen.} {\bf24} (1991), 3557--3574.

\bibitem{levai94} L\'evai G., Solvable potentials associated with su(1,1) algebras: a systematic study,
{\it J.\ Phys.\ A: Math.\ Gen.} {\bf 27} (1994), 3809--3828.

\bibitem{cooper} Cooper F., Khare A., Sukhatme U., Supersymmetry and quantum mechanics, {\it Phys.\
Rep.} {\bf 251} (1995), 267--385, \href{http://arxiv.org/abs/hep-th/9405029}{hep-th/9405029}.

\bibitem{bagchi} Bagchi B., Supersymmetry in quantum and classical mechanics, Chapman and
Hall/CRC, Boca Raton, FL, 2000.

\bibitem{chen} Chen G., Chen Z., Exact solutions of the position-dependent mass Schr\"odinger equation
in $D$ dimensions, {\it Phys.\ Lett.\ A} {\bf 331} (2004), 312--315.

\bibitem{dong} Dong S.-H., Lozada-Cassou M., Exact solutions of the Schr\"odinger equation with the
position-dependent mass for a hard-core potential, {\it Phys.\ Lett.\ A} {\bf 337} (2005), 313--320.

\bibitem{mustafa06a} Mustafa O., Mazharimousavi S.H., $d$-dimensional generalization of the point
canonical transformation for a quantum particle with position-dependent mass, {\it J.\ Phys.\ A:
Math.\ Gen.} {\bf 39} (2006), 10537--10547, \mbox{\href{http://arxiv.org/abs/math-ph/0602044}{math-ph/0602044}}.

\bibitem{mustafa06b} Mustafa O., Mazharimousavi S.H., Quantum particles trapped in a
position-dependent mass barrier; a~$d$-dimensional recipe, {\it Phys.\ Lett.\ A} {\bf 358} (2006),
259--261, \href{http://arxiv.org/abs/quant-ph/0603134}{quant-ph/0603134}.

\bibitem{ju} Ju G.-X., Xiang Y., Ren Z.-Z., The localization of $s$-wave and quantum ef\/fective
potential of a quasi-free particle with position-dependent mass, \href{http://arxiv.org/abs/quant-ph/0601005}{quant-ph/0601005}.

\bibitem{gonul} G\"on\"ul B., Ko\c cak M., Explicit solutions for $N$-dimensional Schr\"odinger
equations with position-dependent mass, {\it J.\ Math.\ Phys.} {\bf 47} (2006), 102101, 6 pages,
\href{http://arxiv.org/abs/quant-ph/0512035}{quant-ph/0512035}.

\bibitem{olendski} Olendski O., Mikhailovska L., Bound-state evolution in curved waveguides and
quantum wires, {\it Phys.\ Rev.\ B} {\bf 66} (2002), 035331, 8 pages.

\bibitem{gudmunsson} Gudmundsson V., Tang C.-S., Manolescu A., Bound state with negative binding
energy induced by coherent transport in a two-dimensional quantum wire, {\it Phys.\ Rev.\ B} {\bf 72}
(2005), 153306, 4 pages, \href{http://arxiv.org/abs/cond-mat/0506009}{cond-mat/0506009}.

\bibitem{goldstein} Goldstein H., Classical mechanics, Addison-Wesley, Reading, MA, 1980.

\bibitem{dirac} Dirac P.A.M., The principles of quantum mechanics, Oxford University Press,
Oxford, 1981.

\bibitem{fris} Fri\v s I., Mandrosov V., Smorodinsky Ya.A., Uhlir M., Winternitz P., On higher
symmetries in quantum mechanics, {\it Phys.\ Lett.} {\bf 16} (1965), 354--356.

\bibitem{winternitz} Winternitz P., Smorodinsky Ya.A., Uhlir M., Fri\v s I., Symmetry groups in
classical and quantum mechanics, {\it Sov.\ J.\ Nucl.\ Phys.} {\bf 4} (1967), 444--450.

\bibitem{makarov} Makharov A.A., Smorodinsky Ya.A., Valiev Kh., Winternitz P., A systematic search
for nonrelativistic systems with dynamical symmetries. Part I: the integrals of motion, {\it Nuovo
Cimento A} {\bf 52} (1967), 1061--1084.

\bibitem{hietarinta} Hietarinta J., Direct methods for the search of the second invariant, {\it Phys.\
Rep.} {\bf 147} (1987), 87--154.

\bibitem{granovskii92a} Granovskii Ya.I., Lutzenko I.M., Zhedanov A.S., Mutual integrability, quadratic
algebras, and dynamical symmetry, {\it Ann.\ Physics} {\bf 217} (1992), 1--20.

\bibitem{zhedanov} Zhedanov A.S., ``Hidden symmetry'' of Askey--Wilson polynomials, {\it Theoret. and
Math.\ Phys.} {\bf 89} (1991), 1146--1157.

\bibitem{granovskii92b} Granovskii Ya.I., Zhedanov A.S., Lutsenko I.M., Quadratic algebras and
dynamics in curved spaces. I.~Oscillator, {\it Theoret. and Math.\ Phys.} {\bf 91} (1992), 474--480.

\bibitem{granovskii92c} Granovskii Ya.I., Zhedanov A.S., Lutsenko I.M., Quadratic algebras and
dynamics in curved spaces. II. The Kepler problem, {\it Theoret. and Math.\ Phys.} {\bf 91} (1992),
604--612.

\bibitem{bonatsos} Bonatsos D., Daskaloyannis C., Kokkotas K., Deformed oscillator algebras for
two-dimensional quantum superintegrable systems, {\it Phys.\ Rev.\ A} {\bf 50} (1994), 3700--3709,
\href{http://arxiv.org/abs/hep-th/9309088}{hep-th/9309088}.

\bibitem{daska01} Daskaloyannis C., Quadratic Poisson algebras of two-dimensional classical
superintegrable systems and quadratic associative algebras of quantum superintegrable systems, {\it
J.\ Math.\ Phys.} {\bf 42} (2001), 1100--1119, \href{http://arxiv.org/abs/math-ph/0003017}{math-ph/0003017}.

\bibitem{daska06a} Daskaloyannis C., Ypsilantis K., Unif\/ied treatment and classif\/ication of
superintegrable systems with integrals quadratic in momenta on a two-dimensional manifold, {\it J.\
Math.\ Phys.} {\bf 47} (2006), 042904, 38 pages, \href{http://arxiv.org/abs/math-ph/0412055}{math-ph/0412055}.

\bibitem{daska06b} Daskaloyannis C., Tanoudes Y., Classif\/ication of quantum superintegrable systems
with quadratic integrals on two dimensional manifolds, \href{http://arxiv.org/abs/math-ph/0607058}{math-ph/0607058}.

\bibitem{letourneau} L\'etourneau P., Vinet L., Superintegrable systems: polynomial algebras and
quasi-exactly solvable Hamiltonians, {\it Ann.\ Physics} {\bf 243} (1995), 144--168.

\bibitem{ranada97} Ra\~nada M.F., Superintegrable $n=2$ systems, quadratic constants of motion, and
potentials of Drach, {\it J.~Math.\ Phys.} {\bf 38} (1997), 4165--4178.

\bibitem{ranada99} Ra\~nada M.F., Santander M., Superintegrable systems on the two-dimensional
sphere $S^2$ and the hyperbolic plane $H^2$, {\it J.\ Math.\ Phys.} {\bf 40} (1999), 5026--5057.

\bibitem{tempesta} Tempesta P., Turbiner A.V., Winternitz P., Exact solvability of superintegrable
systems, {\it J.\ Math.\ Phys.} {\bf 42} (2001), 4248--4257, \href{http://arxiv.org/abs/hep-th/0011209}{hep-th/0011209}.

\bibitem{kalnins97} Kalnins E.G., Miller  W.Jr., Pogosyan G.S., Superintegrability on the
two-dimensional hyperboloid, {\it J.\ Math.\ Phys.} {\bf 38} (1997), 5416--5433.

\bibitem{kalnins99} Kalnins E.G., Miller W.Jr., Hakobyan Y.M., Pogosyan G.S., Superintegrability on the
two-dimensional hyperboloid. II, {\it J.\ Math.\ Phys.} {\bf 40} (1999), 2291--2306,
\href{http://arxiv.org/abs/quant-ph/9907037}{quant-ph/9907037}.

\bibitem{kalnins05a} Kalnins E.G., Kress J.M., Miller  W.Jr., Second-order superintegrable systems in
conformally f\/lat spaces. I.~Two-dimensional classical structure theory, {\it J.\ Math.\ Phys.} {\bf 46}
(2005), 053509, 28 pages.

\bibitem{kalnins05b} Kalnins E.G., Kress J.M., Miller  W.Jr., Second-order superintegrable systems in
conformally f\/lat spaces. II.~The classical two-dimensional St\"ackel transform, {\it J.\ Math.\ Phys.}
{\bf 46} (2005), 053510, 15 pages.

\bibitem{kalnins06} Kalnins E.G., Kress J.M., Miller  W.Jr., Second-order superintegrable systems in
conformally f\/lat spaces. V.~Two- and three-dimensional quantum systems, {\it J.\ Math.\ Phys.} {\bf
47} (2006), 093501, 25 pages.

\bibitem{kalnins07} Kalnins E.G., Kress J.M., Miller  W.Jr., Nondegenerate 2D complex Euclidean superintegrable systems and algebraic varieties, {\it J.\ Phys.\ A: Math.\ Theor.} {\bf 40} (2007), 3399--3411.

\bibitem{cq94} Quesne C., Generalized deformed parafermions, nonlinear deformations of so(3) and
exactly solvable potentials, {\it Phys.\ Lett.\ A} {\bf 193} (1994), 245--250.

\bibitem{daska91} Daskaloyannis C., Generalized deformed oscillator and nonlinear algebras, {\it J.\
Phys.\ A: Math.\ Gen.} {\bf 24} (1991), L789--L794.

\bibitem{roy} Roy B., Roy P., Ef\/fective mass Schr\"odinger equation and nonlinear algebras, {\it Phys.\
Lett.\ A} {\bf 340} (2005), 70--73.

\bibitem{vonroos} von Roos O., Position-dependent ef\/fective masses in semiconductor theory, {\it
Phys.\ Rev.\ B} {\bf 27} (1983), 7547--7552.

\end{thebibliography}
\end{document}